\documentclass[11pt,a4paper,twoside]{article}

\usepackage{graphicx}
\usepackage{fancyhdr,calc,cite,float}
\usepackage{epsfig}
\usepackage[centertags]{amsmath}
\usepackage[mathscr]{eucal}
\usepackage[T1]{fontenc}
\usepackage{amssymb}

\title{\bf Simultaneous loss and excitation 
of projectile electrons in relativistic collisions 
of U$^{90+}$(1s$^2$) ions with atoms } 

\author{\bf B. Najjari and A.B. Voitkiv  \\ 
Max-Planck-Institut f\"ur Kernphysik \\ 
Saupfercheckweg 1, 69117 Heidelberg, Germany }

\date{\today}  

\begin{document}

\maketitle

\begin{abstract} 

We study relativistic collisions between helium-like uranium ions 
initially in the ground state and atoms in which, in a single 
collision event, one of the electrons 
of the ion is emitted and the other is transferred into 
an excited state of the residual hydrogen-like ion. We consider 
this two-electron process at not very high impact energies,  
where the action of the atom on the electrons of the ion 
can be well approximated as occurring solely  
due to the interaction with the nucleus of the atom 
and, hence, the process can be regarded as a four-body problem. 
Using the independent electron model we show that 
a very substantial improvement in the calculated cross sections 
is obtained if, instead of the first order approximation,   
the relativistic symmetric eikonal and 
continuum-distorted-wave-eikonal-initial-state 
models are employed to describe the single-electron 
probabilities for the excitation and loss, respectively.  

\end{abstract}



\section{Introduction} 

The projectile-electron excitation and loss 
occurring in collisions of projectile-ions 
with atomic targets at non-relativistic 
impact energies have been extensively studied 
for long time (for reviews see \cite{MMM}). 
Starting with pioneering articles 
of Bates and Griffing \cite{B-G}   
the most of the theoretical studies of 
these processes have been based on the first order
perturbation theory in the projectile-target 
interaction.  

During the last two decades there has been accumulated 
a very substantial amount of experimental 
data on the electron loss from highly charged ionic 
projectiles moving with relativistic velocities 
and colliding with solid and gaseous targets.  
In particular, various projectiles with atomic numbers 
belonging to the interval $54$-$92$ were used in the experiments. 
The experiments have also covered the very broad 
interval of the projectile impact energies 
ranging from rather low relativistic values 
($0.1$-$0.2$ GeV/u, see e.g. \cite{standf}-\cite{low-middle-3})   
to extreme relativistic ones ($10$ and $160$ GeV/u 
\cite{high-1}-\cite{high-3})    
where the velocity $v$ of the projectile 
differs just fractionally from the speed 
of light $c$ in vacuum  ($ c \approx 137 $ a.u.).  
Besides, a very substantial progress has been 
also reached in the theoretical descriptions 
of the projectile-electron  
transitions occurring 
in relativistic ion-atom collisions 
\cite{theory-1}-\cite{Voit-review}.  

If an ion initially carries several 
electrons, then more than one electron of the ion 
can be simultaneously emitted and/or excited  
in a collision with a neutral atom.  
The explorations of relativistic collisions involving 
simultaneous transitions of two and more projectile 
electrons had began about twenty years ago \cite{standf}. 
However, careful theoretical studies 
of such collisions have not yet been performed. 

Indeed, the theoretical research on this 
topic has, to our knowledge, been restricted 
to a combination of the independent electron model 
with the (simplified) first order theory to calculate 
the probabilities for single-electron transitions 
(see e.g. \cite{E-M} and references therein).   
However, in collisions with heavy atoms the field exerted by 
the atoms on the electrons of the projectile may be 
so strong that attempts to apply the first order approximation  
fail dramatically even for one-electron processes 
yielding results which exceed experimental cross sections 
up to by an order of magnitude \cite{vn-ionization}.       
 
In \cite{thomas} the simultaneous excitation 
and loss of the projectile electrons was investigated 
experimentally for $223.2$ MeV/u U$^{90+}$ ions, 
which are initially in the ground state, impinging 
on rarefied gas targets of argon, krypton and xenon. 
In the collision between the projectile-ion 
and the target-atom one 
of the two electrons of U$^{90+}$ was ejected 
and the other was simultaneously excited into 
the $L$-shell states of U$^{91+}$. Such a process 
represents one of the simplest and basic processes 
which can occur with projectiles having initially 
more than one electron. 

In addition to experimental data the paper \cite{thomas} 
also contains theoretical results. However, 
a satisfactory agreement between these results and 
the experimental data was reached only for collisions 
with argon while in collisions with xenon 
the differences between the theory and experiment  
already reached a factor of $2$-$3$.  

The present article is an attempt to improve 
a theoretical description of the simultaneous 
loss-excitation of the projectile electrons 
in relativistic collisions between very heavy 
helium-like ions and atoms with large atomic numbers. 
This attempt is based on the application 
of relativistic distorted-wave models 
to calculate the single-electron transition 
probabilities. Atomic units are used throughout except where 
otherwise stated. 

\section{Preliminary Remarks}   

Collisions between an helium-like ion 
and a multi-electron or many-electron atom 
are an important and interesting case 
of the quantum many-body problem.  
At present a rigorous and comprehensive treatment of such 
collisions is out of reach.   
In this paper, in order to treat the process of 
the electron loss from a very heavy helium-like ion 
in a collision with a many-electron atom  
in which the corresponding hydrogen-like ion 
is produced in an excited state, we shall  
use a simplified picture of the collision 
consisting of the following main steps. 
 
First, the electrons of the ion 
are regarded as the only particles having dynamical degrees 
of freedom which are described by a quantum treatment. 
Second, the nuclei of the ion and of the atom 
are treated as classical particles which move along given 
(straight-line) trajectories and represent just the sources 
of the external electromagnetic field acting on  
the electrons of the ion. Third, at not very high 
collision energies the excitation and loss of the 
$K$-shell electrons of very heavy ions occur   
at so small impact parameters 
that the electrons of the atom cannot 
effectively screen the nucleus of the atom. 
This is the case already for single-electron 
transitions (see e.g.  \cite{vn-ionization}, 
\cite{vnu-excitation}).   
Since compared to collisions resulting 
in single-electron transitions collisions inducing 
double electron transitions are characterized 
by even smaller impact parameters, the influence 
of the atomic electrons on the simultaneous loss-excitation 
can safely be neglected 
\footnote{ In collisions with many-electron atoms 
the so called antiscreening atomic mode 
is of minor importance for 
electron transitions in highly charged projectiles. 
Moreover, at those collision energies, which we shall 
consider in this article, the effective energy threshold 
for the antiscreening mode will not yet be reached. 
The latter makes this mode unimportant for 
the transitions of the projectile electrons  
also in collisions with a few-electron atoms.}.    

There are essentially two qualitatively 
different possibilities 
to get the simultaneous two-electron transitions. 
The first is that the atom 
in the collision effectively interacts 
with only one electron inducing its transition 
while the other electron 
undergoes a transition due to the 
electron-electron-correlations in the
ion and/or due to rearrangement
in the final state of the ion.  
These processes are called the two-step-1  
and shake-off, respectively (see e.g. \cite{MMM}).   

The other possibility is that the field of the atom 
has enough power to interact simultaneously 
with each of the two electrons 
and to become the main driving force for both electrons 
to undergo transitions. In case when the interaction 
with the atomic nucleus involves the exchange of only 
two virtual photons (one per electron) 
such a process is often referred 
to as the two-step-2 process (see e.g. \cite{MMM}).

It was shown in \cite{we-ratio} that, provided 
the condition $\frac{Z_p Z_t}{v} > 0.4$ is fulfilled  
(where $Z_p$ and $Z_t$ are the charges of the ionic and atomic 
nuclei, respectively, and $v$ the collision velocity),  
two-electron transitions in a heavy helium-like ion 
occurring in collisions with an atom are governed mainly by 
the 'independent' interactions between the nucleus 
of the atom and each of the electrons of the ion. 
This condition, in particular, was very well 
fulfilled in the experiment \cite{thomas}. 
It will be also fulfilled for the collision systems 
considered in this article. 

When the electron transitions are governed by 
the independent interactions the application 
of the independent electron model 
often yields reasonable results. 
According to this model the cross section 
for the simultaneous loss-excitation is evaluated as 
\begin{eqnarray} 
\sigma = 2 \pi \int_0^{\infty} db b P(b),  
\label{ind-1} 
\end{eqnarray} 
where the probability $P(b)$ for 
the two-electron process is given by  
\begin{eqnarray} 
P(b) = 2 \, P_{exc}(b) \, P_{loss}(b).   
\label{ind-2} 
\end{eqnarray} 
Here, $P_{exc}(b)$ and $P_{loss}(b)$ 
are the single-electron excitation 
and loss probabilities, respectively, 
in a collision with a given value of 
the impact parameter $b$. 

Thus, within the simplified picture of 
the ion-atom collision, the theoretical treatment 
of the process of the simultaneous loss and excitation 
has been reduced to finding the electron transition probabilities 
within the three-body problem in which a relativistic electron 
is moving in external electromagnetic fields 
generated by the nuclei of the ion and atom. The latter ones 
are regarded as point-like particles moving 
along given classical trajectories.  

\section{ Single-electron transition amplitudes }

The basic assumption underlying our approach 
to obtain the probabilities 
for single-electron transitions is that the motion 
of the electron in its initial and final states 
is determined mainly by its interaction with 
the field of the ionic nucleus. This interaction, 
therefore, should be treated as accurate 
as possible whereas the interaction of the electron 
with the nucleus of the atom is supposed to be 
less important and, hence, can be taken into account 
in an approximate way. For calculations of the excitation and 
loss probabilities such an 'asymmetric' approach 
seems to be certainly reasonable as long as 
the charge $Z_p$ of the ionic nucleus substantially 
exceeds the charge $Z_t$ of the atomic nucleus 
and the collision velocity is not too low.  

It is convenient to treat the electron transition  
using a reference frame $K$ in which 
the nucleus of the ion is at rest. We take 
the position of this nucleus as the origin 
and assume that in the frame $K$ the nucleus 
of the atom moves along a straight-line 
classical trajectory 
${\bf R}(t)={\bf b} + {\bf v} t$, 
where ${\bf b}=(b_x,b_y,0)$ is 
the impact parameter, ${\bf v}=(0,0,v)$ 
is the collision velocity and $t$ is the time.   

The Dirac equation for the electron of the ion reads 
\begin{eqnarray}
i \frac{ \partial \Psi }{\partial t} = %
\left( \hat{H}_0 + \hat{W}(t) \right) \Psi.      
\label{e1} 
\end{eqnarray}
Here 
\begin{eqnarray}  
\hat{H}_0 = c {\mbox{\boldmath $\alpha$} } %
\cdot \hat{\bf p} - \frac{Z_p}{r} + \beta c^2     
\label{e2}
\end{eqnarray}
is the electronic Hamiltonian for 
the undistorted ion and   
\begin{eqnarray}
\hat{W}(t) = - \Phi({\bf r},t) + {\mbox{\boldmath $\alpha$} } %
\cdot {\bf A}({\bf r},t) 
\label{e3}
\end{eqnarray}
is the interaction between the electron of the ion 
and the nucleus of the atom, 
where $\Phi$ and ${\bf A}$ are the scalar and vector 
potentials of the electromagnetic field generated 
by the atomic nucleus in the frame $K$.  
In the above equations $\hat{\bf p}$ is 
the electron momentum operator,  
${\mbox{\boldmath $\alpha$}}=(\alpha_x, \alpha_y, \alpha_z)$ 
and $\beta$ are the Dirac matrices, 
$Z_p$ is the charge of the nucleus 
of the ion and ${\bf r}=(x,y,z)$ are the 
electron coordinates with respect to this nucleus.   

The potentials $\Phi$ and ${\bf A} $ of the 
moving atomic nucleus will be taken 
in the Lienard-Wiechert form 
(see e.g. \cite{Jack} )    
\begin{eqnarray}
\Phi({\bf r},t) &=& \frac{ \gamma Z_t}{s} 
\nonumber \\   
{\bf A}({\bf r},t) &=& \frac{\bf v}{c} \Phi({\bf r},t),    
\label{lien-wiech}
\end{eqnarray}
where $Z_t$ is the charge of the nucleus of the atom, 
${\bf s}=(s_x,s_y,s_z)=(x-b_x,y-b_y,\gamma(z-v t))$ and 
$\gamma=1/\sqrt{1-v^2/c^2}$ is the collisional Lorentz factor, 

The prior form of the semi-classical 
transition amplitude is given by 
\begin{eqnarray}
a_{fi}({\bf b}) = - i \int_{- \infty}^{+ \infty} dt %
\langle \Psi^{(-)}(t) \mid %
\left( \hat{H}_0 + \hat{W}(t) %
 - i \partial/\partial t \right) \chi_i(t) \rangle. 
\label{e4}
\end{eqnarray}
In (\ref{e4}) $\Psi^{(-)}(t)$ is the solution of the full
Dirac equation (\ref{e1}) 
and $\chi_i(t)$ is the solution of
\begin{eqnarray}
i \frac{ \partial \chi_i }{\partial t} = %
\left( \hat{H}_0 + V(t) \right) \chi_i, 
\label{e5}
\end{eqnarray}
where $ V(t) $ is a distortion potential.
The initial condition reads $\chi_i(t \to -\infty) \to %
\psi_i \exp(-i\varepsilon_i t ) $, where 
$ \psi_i$ is the (undistorted) initial state 
of the electron of the ion with
an energy $ \varepsilon_i$.   

In what follows it will be much more convenient 
to obtain first the transition amplitude  
in the momentum space, $S_{fi}( {\bf Q} )$, 
which is related to the amplitude (\ref{e4}) 
by the two-dimensional Fourier transformation  
\begin{eqnarray}
S_{fi}( {\bf Q} ) = \frac{1}{2 \pi} %
\int d^2 {\bf b} \, a_{fi}({\bf b}) \, %
\exp(i {\bf Q} \cdot {\bf b} ),   
\label{e6} 
\end{eqnarray} 
and then to use the inverse transformation,  
\begin{eqnarray}
a_{fi}( {\bf b} ) = \frac{1}{2 \pi} %
\int d^2 {\bf Q} \, S_{fi}({\bf Q}) \, %
\exp(- i {\bf Q} \cdot {\bf b} ),   
\label{e7} 
\end{eqnarray} 
in order to get the transition amplitude 
in the impact parameter space. 

\subsection{First order approximation} 

The first order transition amplitude is obtained 
by replacing the states $\chi_i(t)$ and $\Psi^{(-)}(t)$ 
by the undistorted initial, $\psi_i$, 
and final, $\psi_f$, states 
of the electron in the ion: 
\begin{eqnarray} 
\chi_i(t)&=&\psi_i({\bf r}) exp(-i \varepsilon_i t) 
\nonumber \\ 
\Psi^{(-)}(t)&=&\psi_f({\bf r}) exp(-i \varepsilon_f t),  
\label{1order-1}
\end{eqnarray}   
where $\varepsilon_f$ is the energy of 
the electron in the final state of the ion. 
Using (\ref{e3}), (\ref{lien-wiech}), (\ref{e4}), (\ref{e6})  
and (\ref{1order-1}), the first order amplitude in the momentum 
space is obtained to be 
\begin{eqnarray}
S_{fi}^{(1)}( {\bf Q} ) = \frac{2 i Z_t}{v} %
\frac{1}{q'^2} \left( %
\langle \psi_f \mid \exp(i {\bf q} %
\cdot {\bf r}) \mid \psi_i \rangle %
- \frac{v}{c} %
\langle \psi_f \mid \exp(i {\bf q} %
\cdot {\bf r}) \alpha_z \mid \psi_i \rangle %
\right),      
\label{1order-2} 
\end{eqnarray} 
where the momentum ${\bf q}$, which is transferred 
to the electron of the ion in the collision, 
is given by  
\begin{eqnarray}
{\bf q} &=& \left(q_x,q_y,q_z \right)= %
\left( {\bf Q}; q_z \right) %
\nonumber \\ 
q_z &=& \frac{\varepsilon_f-\varepsilon_i}{v}   
\label{momentum-1} 
\end{eqnarray} 
and 
\begin{eqnarray}
{\bf q}' &=& \left(q_x,q_y,q_z/\gamma \right) =%
\left( {\bf Q}; q_z/\gamma \right).  
\label{momentum-2}  
\end{eqnarray}  

\subsection{Symmetric eikonal approximation} 
 
In an attempt to obtain a better description of 
the interaction between the electron of the ion 
and the nucleus of the atom we now approximate 
the initial and final states in (\ref{e4}) 
by $\chi_i$ and $\chi_f$, respectively, where 
\begin{eqnarray}
\chi_i(t) &=& \psi_i({\bf r})  %
\times (v s + {\bf v} \cdot {\bf s} )^{ - i \nu_t } %
\times \exp(- i \varepsilon_i t) %
\nonumber \\
\chi_f(t) &=& \psi_f({\bf r})  %
\times (v s - {\bf v} \cdot {\bf s} )^{ i \nu_t } %
\times \exp(- i \varepsilon_f t)  %
\label{sea-1}
\end{eqnarray}
with $\nu_t= Z_t/v$. The semiclassical transition 
amplitude is then given by  
\begin{eqnarray}
a_{fi}^{eik}({\bf b}) = - i \int_{- \infty}^{+ \infty} dt %
\langle \chi^{\dagger}_f(t) \mid \hat{W}_d(t) \chi_i(t) \rangle,   
\label{sea-2}
\end{eqnarray}
where the action
of the distortion interaction 
$\hat{W}_d(t)=\hat{H} - i \frac{ \partial }{ \partial t}$
on the initial state is defined according to  
\begin{eqnarray}
\hat{W}_d \chi_i &=& - \frac{c Z_t}{v}%
(v s + {\bf v} \cdot {\bf s})^{- i \nu_t} 
\nonumber \\ 
&& \times \left( \frac{s_x \alpha_x + s_y \alpha_y}{s+s_z} + 
\frac{\alpha_z}{\gamma} \right) \psi_i \exp(- i \varepsilon_i t ). 
\label{sea-3}
\end{eqnarray}
Using Eqs.(\ref{sea-1})-(\ref{sea-3}) and (\ref{e6}) 
it can be shown \cite{vnu-excitation} that 
the symmetric eikonal amplitude 
in the momentum space reads 
\begin{eqnarray}
S_{fi}^{eik}( {\bf Q} ) &=& %
\frac{2 i Z_t c }{ v^2 } \, %
\frac{ 1 }{q'^2 q_z} \, %
\left(\frac{q'}{2}\right)^{2i\nu_t} %
\, \Gamma^2(1-i \nu_t) \times %
\nonumber \\ 
&& \left( (1-i \nu_t) \, %
\left._2F_1\left(1-i\nu_t,i\nu_t;2; Q^2/q'^2 \right) \right.  %
\langle \psi_f \mid \exp(i {\bf q} \cdot {\bf r}) %
(q_x \alpha_x + q_y \alpha_y) %
\mid \psi_i \rangle \right. %
\nonumber \\ 
&& \left. + 
\left._2F_1\left(1-i\nu_t,i\nu_t;1; Q^2/q'^2 \right) \right. %
\, \frac{1}{\gamma^2} \, %
\langle \psi_f \mid \exp(i {\bf q} \cdot {\bf r}) q_z \alpha_z %
\mid \psi_i \rangle \right), %
\label{sea-4} 
\end{eqnarray}
where $\Gamma$ and 
$ \left._2F_1 \right.$  
are the gamma and hypergeometric functions, 
respectively (see e.g. \cite{Ab-St}). 
Note that, provided $\psi_i$ and $\psi_f$ are 
exact states of the undistorted Hamiltonian 
of the ion given by Eq.(\ref{e2}), 
in the limit of weak perturbation  
$\nu_t \to 0$, the amplitude  
(\ref{sea-4}) goes over into  
the first order amplitude (\ref{1order-2}). 

With the help of Eq.(\ref{e7}) the amplitude 
(\ref{sea-4}) is converted into the impact 
parameter space and the corresponding eikonal 
amplitude $a_{fi}^{sea}({\bf b})$ will be used 
to calculate the probability $P_{exc}(b)$ 
for single-electron excitation. 

\subsection{Continuum-distorted-wave-eikonal-initial-state approximation} 

Within the symmetric eikonal approximation, 
in order to reflect the influence of the nucleus 
of the atom on the electron, its initial and final states 
are distorted similarly. This might be a suitable choice 
if both initial and final electron states are bound states. 
When the final electron state is a continuum state 
a more sophisticated distortion factor is in general necessary.  
Therefore, in an attempt to obtain a better  
(compared to the first order theory) 
description of the electron continuum states, 
we now approximate the initial   
and final states in (\ref{e4}) in the spirit of  
the continuum-distorted-wave-eikonal-initial-state   
(CDW-EIS) approach \cite{croth-mccann}, \cite{deco-gruen} by    
\begin{eqnarray}
\chi^{eik}_i(t) &=& \psi_i({\bf r}) \exp(- i \varepsilon_i t) %
\times (v s + {\bf v} \cdot {\bf s} )^{ - i \nu_t } %
\nonumber \\
\chi^{cdw}_f(t) &=& \psi_f({\bf r}) \exp(- i \varepsilon_f t) %
\times \Gamma(1 + i \eta_t) \exp(\pi \eta_t/2) %
\left._1F_1\left(-i \eta_t,1,-i(p' s + {\bf p}' %
\cdot {\bf s}) \right) \right. . 
\label{cdw-1}
\end{eqnarray}
Here, $\eta_t = Z_t/v'_e$ where $v'_e$ 
is the electron velocity with respect 
to the nucleus of the atom as given  
in the rest frame of this nucleus, ${\bf p}'$ 
($p'=|{\bf p}'|$) is the kinetic momentum 
of the electron in the rest frame 
of the atomic nucleus and $ \left._1F_1 \right. $  
is the confluent hypergeometric function 
(see e.g. \cite{Ab-St}). 

Using Eqs.(\ref{e4}), (\ref{e6}), (\ref{sea-3}) 
and (\ref{cdw-1}) we obtain that the CDW-EIS transition amplitude 
in the momentum space is given by 
(see \cite{vn-ionization})   
\begin{eqnarray}
S_{fi}^{cdw-eis}( {\bf Q} ) = \frac{2i Z_t c }{\gamma v } %
\left(\frac{A}{C} \right)^{i\nu_t} %
\left( \frac{A+B}{A}\right)^{-i\eta_t}
\langle \psi_f|J_x \alpha_x + J_y \alpha_y + %
J_z \alpha_z |\psi_i \rangle,     
\label{cdw-2} 
\end{eqnarray}     
where 
\begin{eqnarray}
J_{x (y)} &=& \frac{ \Gamma(-i \nu_t) }{C} %
\left( \Omega_{x (y)} \left._2F_1 \left(i\nu_t,i\eta_t,1,Z \right) \right. 
 + \Omega'_{x (y)} \left._2F_1 \left(i\nu_t+1,i\eta_t+1,2,Z \right) \right. \right),  
\nonumber \\ 
J_z &=& \frac{ \Gamma(1-i \nu_t) }{A \gamma v}  
\left._2F_1 \left(i\nu_t,i\eta_t,1,Z \right) \right., 
\label{cdw-3}
\end{eqnarray}    
\begin{eqnarray}
A= q'^2, \, \, B = - 2 i {\bf q}'\cdot {\bf p}, \, \,   
C = - 2 i q'_z v, \, \,  D = 2i v (p_z -p), \, \,   
Z= \frac{B C-A D}{C(A+B)} 
\label{cdw-4}
\end{eqnarray}    
and 
\begin{eqnarray}
\Omega_{x (y)} &=& (\nu_t+\eta_t) \frac{\partial \ln(A)}{\partial q_{x (y)}}  
-\eta_t \frac{\partial \ln(A+B)}{\partial q_{x (y)}} + \eta_t \frac{Z}{1-Z}
\frac{\partial \ln(Z)}{\partial q_{x (y)}}    
\nonumber \\ 
\Omega'_{x (y)} &=& i \eta_t Z \left( \Omega_{x (y)} + \frac{\nu_t}{1-Z}  %
\frac{\partial \ln(Z)}{\partial q_{x (y)}} \right).
\label{cdw-5}
\end{eqnarray} 
Note that in the limit of weak perturbation, 
$\nu_t \to 0$ and $\eta_t \to 0$, the amplitude  
(\ref{cdw-2}) coincides with the first order amplitude 
(\ref{1order-2}) provided $\psi_i$ and $\psi_f$ are 
exact states of the undistorted Hamiltonian 
of the ion $\hat{H}_0$.  

Using Eqs.(\ref{cdw-2})-(\ref{cdw-5}) 
and (\ref{e7}) we can calculate 
numerically the corresponding CDW-EIS amplitude 
$a_{fi}^{cdw-eis}( {\bf b} )$  
in the impact parameter space and evaluate    
the total probability $P_{loss}(b)$ 
for single-electron loss. 
 
\section{ Results and discussion }

\begin{figure}[t] 
\vspace{0.25cm}
\begin{center}
\includegraphics[width=0.77\textwidth]{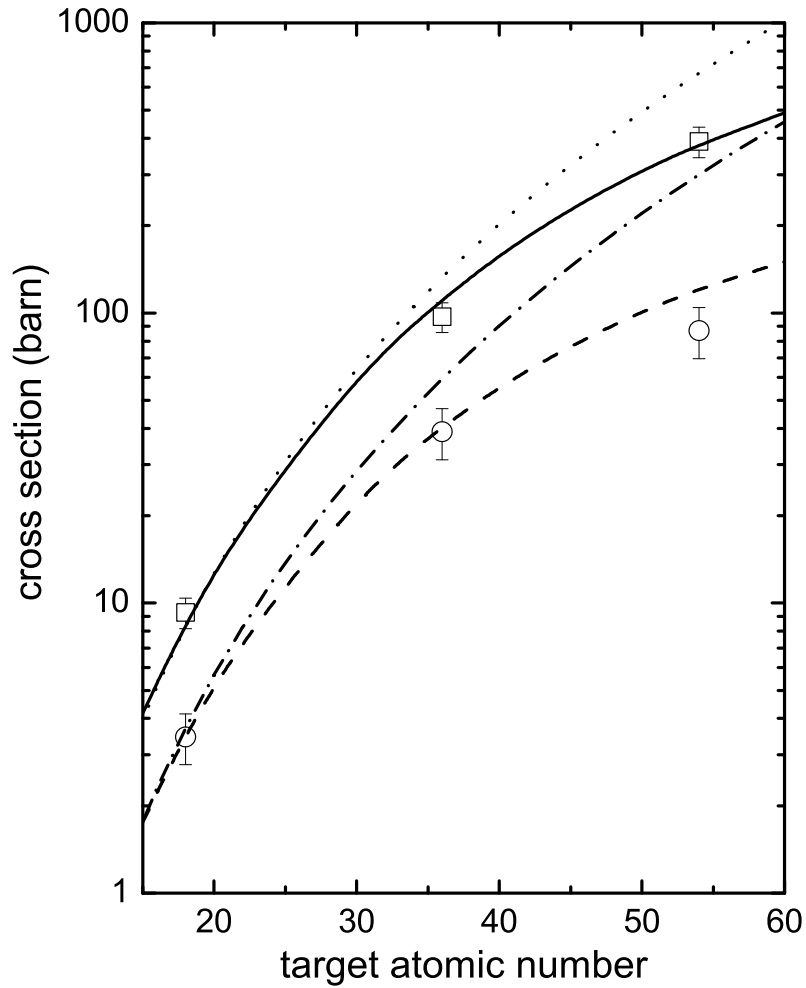}
\end{center}
\caption{ Total cross sections for the reactions 
$223.2$ U$^{90+}$(1s$^2$) $+ Z_t$ $\to$ 
U$^{91+}$(n=2, j) + e$^-$ +..., 
where $j=1/2$ and $j=3/2$ are the angular momentum 
of the $L$-shell states of the hydrogen-like uranium ion.  
The cross sections are given as a function of 
the atomic number $Z_t$ of the target atom. 
Circles and squares with the corresponding error bars 
are experimental data for the $j=3/2$ and $j=1/2$ cases, 
respectively, reported in \cite{thomas} 
for collisions with argon, krypton and xenon gas targets. 
Dot ($j=1/2$) and dash-dot ($j=3/2$) curves show 
the cross sections calculated with 
the single-electron transition probabilities 
obtained in the first order of perturbation 
theory in the interaction between the electron and 
the nucleus of the atom. 
Solid ($j=1/2$) and dash ($j=3/2$) curves 
display theoretical results obtained  
using the symmetric eikonal approximation 
to estimate the excitation probability 
and the CDW-EIS approximation 
to calculate the loss probability. }
\label{exc-ion}
\end{figure}

In figure \ref{exc-ion} we show 
experimental data for 
the cross sections for the simultaneous 
loss-excitation of the projectile electrons.  
These cross sections were measured in \cite{thomas}    
for collisions of $223$ MeV/u U$^{90+}(1s^2)$ with 
gas targets of argon, krypton and xenon for the cases 
when the internal states of the hydrogen-like projectile ion 
formed in these collision were characterized by 
the principal quantum number $n=2$ and the total angular momentum 
$j=1/2$ or $j=3/2$. Figure \ref{exc-ion} also contains 
our theoretical results.    

As was already mentioned, at this relatively low 
impact energy the screening effect of the atomic electrons 
has a negligible impact on the loss-excitation process. 
Besides, since the effective energy threshold 
for the antiscreening target mode 
in this process is about $420$ MeV/u, 
the antiscreening effect of the atomic electrons 
in the collisions under consideration is also very weak. 
Thus, the presence of the atomic electrons 
can indeed be ignored, 
the single-electron transition probabilities 
can be obtained within the scope of 
an effectively three-body problem and we may apply 
the three-body approximations discussed 
in the previous section. 

Two sets of theoretical results are presented  
in figure \ref{exc-ion}. One of them 
was obtained using the first order perturbation theory 
to calculate the single-electron excitation 
and loss probabilities 
\footnote{ The authors 
of \cite{thomas} had also presented 
theoretical results obtained 
using the first order perturbation theory 
to calculate the excitation and loss probabilities. 
The main difference between their 
and our first order results lies 
in the following. In \cite{thomas} 
the loss probability was calculated using 
the computer code of \cite{trautmann} in which the interaction 
between the electron of the ion and the nucleus of the atom 
is approximated by the unretarded (nonrelativistic) 
Coulomb interaction. In our calculations 
the fully relativistic form 
of this interaction has been used.}. 
In the other approach the excitation probability 
$P_{exc}(b)$ was evaluated using 
the symmetric eikonal approximation and 
the loss probability $P_{loss}(b)$ was calculated 
in the CDW-EIS approximation.  
In both the first order and distorted-wave calculations 
the states $\psi_i$ and $\psi_f$ were approximated by 
relativistic bound and continuum states of a hydrogen-like 
ion with a nuclear charge $Z_p=92$. 

It is seen in figure \ref{exc-ion} that in collisions with 
atomic targets having not very large atomic numbers, for 
which one has $Z_t/v \ll 1$, the two theoretical approaches  
yield very close cross section values.  
When the ratio $Z_t/v $ increases,   
the differences between the results of 
the first order approximation and 
the distorted-wave calculations rapidly grow. 

A comparison with the experimental data clearly 
shows a strong failure of the first order calculations. 
According to the latter ones the cross section 
for the two-electron transitions should depend 
on the target atomic number $Z_t$ as $\sim Z_t^4 $. 
However, the experimental points 
indicate that 
there are substantial deviations from the 
$Z_t^4$-dependence which increase  
with increasing $Z_t$. For instance, 
for collisions with xenon the first order results   
overestimates the experimental data 
already by a factor of $1.7$ ($j=1/2$) and $3$ ($j=3/2$).   

Compared to the first order calculations, 
the calculations employing the distorted wave models 
predict a much slower increase 
in the cross sections when the atomic number 
of the target becomes sufficiently large.  
Besides, the magnitude of the deviations 
between the results, obtained with the first 
order and the distorted wave 
calculations, depends on which excited 
states ($j=1/2$ or $j=3/2$) 
are occupied by the remaining electron 
and are accumulating at a different pace 
for different final electron states.   

Comparing the experimental data with 
the results of the distorted wave calculations  
we see that these calculations 
turned out to be much more successful than 
the first order ones. Nevertheless, 
a certain disagreement between the experiment 
and theory remains also in the case of 
the distorted wave calculations.  

\begin{figure}[t]
\vspace{0.25cm} 
\begin{center}
\includegraphics[width=0.77\textwidth]{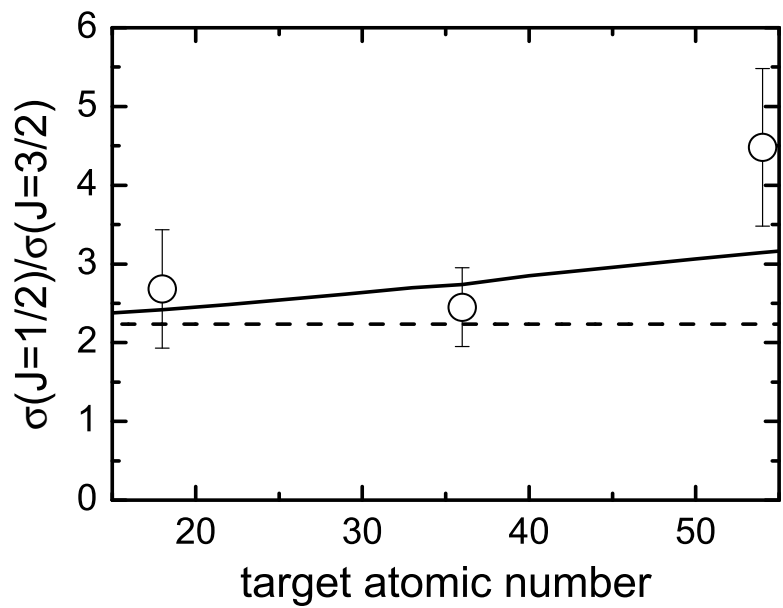}
\end{center}
\caption{ \footnotesize{ The ratio 
between the cross sections $\sigma(j=1/2)$  
and $\sigma(j=3/2)$ for the simultaneous 
loss-excitation processes occurring 
in 223 MeV/u U$^{90+}(1s^2)$ ions  
in which after the collision  
the corresponding hydrogen-like ions  
occupy the $n=2$ states with the total 
angular momenta $j=1/2$ and $j=3/2$, respectively. 
The ratio is given as a function 
of the atomic number of the target.  
Dash line: results of the first order calculation. 
Solid curve: results obtained with the distorted-wave 
approach. Circles with error bars are  
experimental data from \cite{thomas}. } } 
\label{ratio}  
\end{figure} 

The latter is also seen in figure \ref{ratio} 
where we compare results for the ratio 
$\sigma(n=2,j=1/2)/\sigma(n=2,j=3/2)$. 
Here, $\sigma(n=2,j=1/2)$ and $\sigma(n=2,j=3/2)$ 
are the cross sections for the loss-excitation 
processes in which the electron remaining bound 
has been excited into the states with $n=2,j=1/2$ 
and $n=2,j=3/2$, respectively. 

\section{ Conclusions }

We have calculated the total cross sections for 
the simultaneous loss and excitation of the projectile 
electrons in collisions of $223$ MeV/u helium-like uranium ions 
with various atomic targets with atomic numbers 
ranging between $15$ and $60$. 
In these collisions the effect 
of the atomic electrons on the simultaneous 
loss-excitation is weak and can be neglected. 
At the same time, in such collisions  
the ratio $Z_t/v$ can reach values which are 
not much less than $1$ meaning  
that the field of the atomic nucleus 
acting on the electrons of the ion 
in these collisions can effectively 
be quite strong. 

Two different theoretical approaches 
were used in our study. Both of them  
are based on the independent electron approximation. 
Within the framework of the latter the treatment 
of the process of the simultaneous loss-excitation 
is essentially reduced to two steps: 
i) obtaining the transition probabilities 
as a function of the impact parameter and 
ii) making use of these probabilities 
to calculate the cross sections. 

Within the first approach 
the single-electron probabilities for the loss and excitation 
were evaluated in the first order 
approximation in the interaction 
between the electron of the projectile and the 
nucleus of the target. In the treatment of 
the simultaneous loss-excitation 
occurring in collisions with rather heavy  
atoms this approach was shown to strongly fail.  

Within the second approach the probability for 
the single-electron excitation  
was calculated in the symmetric 
eikonal approximation and the probability 
for the electron loss was evaluated 
using the continuum-distorted-wave-eikonal-initial-state 
model. Compared to the first order approximation, 
the distorted-wave treatments have enabled us 
to reach very substantial improvements 
in the description of the experimental data.  
Nevertheless, slight disagreements between 
the theory and experiment are still remaining 
which indicates that additional theoretical 
(and perhaps experimental) work is necessary.

\end{document}